%Paper: hep-ph/9304214
%From: EZ <ZYNDA%IASSNS.BITNET@pucc.Princeton.EDU>
%Date: Mon, 5 Apr 93 07:44 EST

\input phyzzx
\hoffset=0.2truein
\voffset=0.1truein
\hsize=6truein
\def\TITLEPAGE{\frontpagetrue}
\def\IASSNS#1{\hbox to\hsize{\tenpoint \baselineskip=12pt
        \hfil\vtop{
        \hbox{\strut iassns-hep-92-#1}
}}}

\def\IAS{
        \centerline{School of Natural Science, Institute for Advanced Study}
\centerline{Olden Lane, Princeton, New Jersey 08540}}
\def\TITLE#1{\vskip .5in \centerline{\fourteenpoint #1}}
\def\AUTHOR#1{\vskip .2in \centerline{#1}}

\def\ABSTRACT#1{\vskip .2in \vfil \centerline{\twelvepoint
\bf Abstract}
   #1 \vfil}
\def\ENDTITLEPAGE{\vfil\eject\pageno=1}
\def\comment#1{}
\hfuzz=5pt
\tolerance=10000
\def\Chi{\raise 0.5ex\hbox{$\chi$}}
\TITLEPAGE
\IASSNS{92}
\TITLE{Skyrmions and Semilocal Strings in Cosmology}
\vskip 20pt

\AUTHOR{Katherine Benson\footnote\dagger{Supported in
part by NSF grant PHY91-6210} and Martin Bucher\footnote\ddagger{Supported
in part by the US Department of Energy grant
DE-FG02-90ER40542}}
\vskip 10pt

\IAS

{\baselineskip=14pt\ABSTRACT{
\bigskip
It has been pointed out that cosmic string
solutions can exist in gauge field theories with broken
symmetry even when $\pi _1(G/H)$ is trivial. The stability
of such semilocal defects is not guaranteed by topology
and depends on dynamical considerations. In the literature
it has been tacitly assumed that if stable, such strings
would form in the Early Universe in a manner analogous
to the formation of a network of more robust
topologically-stable strings. In this paper we
find that except for unnaturally small values of
the correlation length, a network of semilocal strings
does not form. Instead, delocalized skyrmionic string
configurations, which expand with the Hubble flow,
dominate.
\bigskip\rightline{April 1993}
}}
\ENDTITLEPAGE
\eject

\def\gtorder{\mathrel{\raise.3ex\hbox{$>$}\mkern-14mu
             \lower0.6ex\hbox{$\sim$}}}
\def\ltorder{\mathrel{\raise.3ex\hbox{$<$}\mkern-14mu
             \lower0.6ex\hbox{$\sim$}}}

\REF\preskill{J. Preskill, ``Vortices and
Monopoles,'' in {\it Architecture of the Fundamental Interactions at Short
Distances}, ed. P. Ramond and R. Stora (North-Holland, Amsterdam,
1987).}

\REF\vilenkin{A. Vilenkin, ``Cosmic Strings and Domain Walls,"
Phys. Rep. {\bf 121,} 263 (1984)}

\REF\book{G. Gibbons, S. Hawking and T. Vachaspati, Eds.,
{\it The Formation and Evolution of Cosmic Strings,}
Cambridge: Cambridge U. Press (1990).}

\REF\abr{A. Abrikosov, ``On the Magnetic Properties of Superconductors
of the Second Group," Soviet Phys. JETP {\bf 5,} 1174 (1957).}

\REF\no{H. Nielsen and P. Oleson, ``Vortex-Line Models for Dual
Systems," {\bf B61,} 45 (1973).}

\REF\vacha{T. Vachaspati and A. Ach\'ucarro, ``Semilocal Cosmic
Strings," Phys. Rev. Lett. {\bf 44,} 3067 (1991).}

\REF\nambu{Y. Nambu, ``Stringlike Configurations in the Weinberg-Salam
Theory," Nucl. Phys. {\bf B130,} 505 (1978).}

\REF\tvtwo{T. Vachaspati, ``Vortex Solutions in the Weinberg-Salam
Model," Phys. Rev. Lett. {\bf 68,} 1977 (1992).}

\REF\hinda{M. Hindmarsh, ``Existence and Stability of Semilocal
Strings," Phys. Rev. Lett. {\bf 68,} 1263 (1992).}

\REF\akpv{A. Ach\'ucarro, K. Kuijken, L. Perivolaropoulos and
T. Vachaspati, ``Dynamical Simulations of Semilocal Strings,"
Harvard-Smithsonian CfA Preprint 3384 (1992).}

\REF\james{M. James, L. Perivolaropoulos and T. Vachaspati,
``Detailed Stability Analysis of Electroweak Strings,"
Tufts Preprint 1992}

\REF\sew{L. Perivolaropoulos and T. Vachaspati, ``On the Stability
of Electroweak Strings,
Harvard-Smithsonian CfA Preprint 3388 (1992).}

\REF\gwg{G. Gibbons, M. Ortiz, F. Ruiz and T. Samols, ``Semilocal
Strings and Monopoles," DAMTP HEP 92-09 (1992).}

\REF\leese{R. Leese and T. Samols, ``Interactions of Semilocal Vortices,"
DAMTP 92/40 (1992).}

\REF\vw{T. Vachaspati and R. Watkins, ``Bound States Can
Stabilize Electroweak Strings," TUPT-92-10 (1992).}

\REF\hol{R. Holman, S. Hsu, T. Vachaspati and R. Watkins,
``Metastable Cosmic Strings in Realistic Models,"
Phys. Rev. {\bf D46,} 5352 (1992).}

\REF\mb{T. Vachaspati and M. Barriola, ``A New Class of Defects,"
Phys. Rev. Lett. {\bf 69,} 1867 (1992).}

\REF\preskillb{J. Preskill, ``Semilocal Defects," CALT-68-1787 (1992).}

\REF\htwo{M. Hindmarsh, ``Semilocal Toplogical Defects,"
DAMTP-HEP-92-24 (1992).}

\REF\bogo{E.B. Bogomol'nyi, ``Stability of Classical Solutions,"
Sov. J. Nucl. Phys. {\bf 24,} 449 (1976).}

\REF\pv{J. Preskill and A. Vilenkin, ``Decay of Metastable Topological
Defects," Harvard Preprint HUTP-92-A018 (1992).}

\REF\kibble{T.W.B. Kibble, ``Topology of Cosmic Domains and Strings,"
J. Phys. {\bf A9,} 1387 (1976).}

\REF\zel{Y. Zel'dovich, ``Cosmological Fluctuations Produced Near a
Singularity," Mon. Not. Astr. Soc. {\bf 192,} 663 (1980).}

\REF\vila{A. Vilenkin, ``Cosmological Density Fluctuations
Produced by Vacuum Strings," Phys. Rev. Lett. {\bf 46,} 1169 (1981),
Erratum: {\bf 46,} 1496 (1981).}

\REF\vv{T. Vachaspati and A. Vilenkin, ``Formation and Evolution
of Cosmic Strings," Phys. Rev. {\bf D30,} 2036 (1984).}

\REF\ktwo{T.W.B. Kibble, ``Evolution of a System of Cosmic Strings,"
Nucl. Phys. {\bf B252,} 227 (1985).}

\REF\shellard{E.P.S. Shellard, ``Cosmic String Interactions,"
Nucl. Phys. {\bf B283,} 624 (1987)}

\REF\matzner{R. Matzner, Comput. Phys. {\bf 2,} 51 (1988).}

\REF\mor{K. Moriarty, E. Meyers and C. Rebbi, ``Interactions
of Cosmic Strings," in F. Acceta and L. Krauss, {\it Cosmic
Strings: The Current Status,} Singapore: World Scientific (1988).}

\chapter{Introduction}

When there is a pattern of symmetry breaking $G\to H$
through the formation of a condensate, there is a
manifold of degenerate vacua that is homeomorphic to the coset
space $G/H.$ When the homotopy groups
$\pi _k(G/H)$
are nontrivial, topologically-stable defects arise in
the theory.
\refmark{\preskill, \vilenkin }
In 3+1 dimensions, domain walls result for
nontrivial $\pi _0(G/H),$ cosmic strings for nontrivial
$\pi _1(G/H),$ monopoles for nontrivial $\pi _2(G/H),$ and
textures for nontrivial $\pi _3(G/H).$
The physics of such defects in quantum field theory has been
studied in much detail.
Topological defects form the basis of several proposed scenarios
for structure formation in the Early Universe. \refmark{\book }

Recently there has been much discussion of embedded defects,
especially embedded strings, which may exist even when $\pi _1(G/H)$
is trivial \refmark{\vacha -\htwo }.
Like their conventional counterparts, embedded strings,
with a Higgs field winding about a line defect carrying magnetic flux,
satisfy the classical field equations.
However, because they arise in theories with trivial $\pi_1(G/H),$
the stability of these embedded solutions does not follow from
topological considerations, but rather depends on dynamical considerations.
Despite their more fragile basis, nontopological strings are
stable against small perturbations for a natural range of parameters.

The simplest model with embedded strings is an extension of the
$U(1)$ abelian Higgs model with the familiar topologically-stable
Abrikosov-Nielsen-Oleson strings \refmark{\abr ,\no }.
In this  extended model, considered by Vachaspati and
Ach\'ucarro \refmark{\vacha }, the complex Higgs singlet is replaced
by a complex Higgs doublet, so that in addition to the gauged
$U(1)$ local symmetry, there is also an $SU(2)$ global symmetry,
and the vacuum manifold has the topology of $S^3.$
The action of $U(1)$ foliates
$S^3$ into gauge orbits, each with the structure of $S^1.$
For the embedded strings---called ``semilocal strings" by
Vachaspati and Ach\'ucarro---the Higgs field is restricted
to one of these gauge orbits. The string is not topologically
stable because the gauge field can unwind by slipping off the
gauge orbit. However, such a perturbation need not
lower the energy
because the scalar gradient energy is increased by the
perturbations. Hindmarsh \refmark{\hinda } examined
the issue of stability in this model and found that
stability is determined by the dimensionless parameter
$\beta = m_H^2/ m_V^2 = \lambda /g^2.$ For $\beta <1$
the solution is stable against small perturbations;
for $\beta >1$ the solution is unstable.

Since semilocal strings are not necessarily stable, the
question arises: Into what can they decay? In the model
described above, there is a competing family of configurations,
considered by both Hindmarsh and Preskill
\refmark{\hinda , \htwo , \preskillb }, dubbed ``skyrmions"
by Preskill. These ``skyrmions" are essentially global textures
in two spatial dimensions---or alternatively, fat delocalized strings.
At very large radii, the Higgs field
winds around a gauge orbit and is compensated by the gauge field
so that the covariant derivative vanishes, just as for the
localized semilocal strings. However, inside the core of
the skyrmion the Higgs field does not vanish as in the
Abrikosov-Nielsen-Oleson solution, but rather remains
on (or very close to) the vacuum manifold. Said another way,
the twist around a gauge orbit at spatial infinity
unwinds in the core of the skyrmion while the Higgs field
essentially remains on the vacuum manifold. The existence
of such skyrmionic configurations follows directly
from the triviality of $\pi _1(G/H).$ Because the Higgs field
need not leave the vacuum manifold to unwind the twist at
infinity, skyrmions can be made arbitrarily large.
For very large skyrmions, the tension is dominated by
scalar gradient energy that cannot be compensated by the gauge
field. The tension of a large skyrmion (that is, $R\gg v^{-1}$)
is of order $v^2,$ where $v$ is the symmetry breaking scale. It is a dynamical
question whether this tension is greater than or less than
the tension of the Abrikosov-Nielsen-Oleson string, which
also is of order $v^2.$

The existence of competing low-energy, delocalized skyrmion
configurations complicates the analysis of the role of semilocal
strings in cosmology. Because the frustration generated by
quenching in a cosmological phase transition
can be resolved either through the formation of
localized semilocal strings or through the formation
of delocalized skyrmionic strings, the usual analysis of string
formation and evolution breaks down, and an understanding of skyrmion
dynamics is crucial to determining the role of semilocal strings
in cosmology.

Both Hindmarsh and Preskill have considered the dynamics of
skyrmions \refmark{\preskillb ,\hinda , \htwo }. Preskill
considered skyrmions in the $CP^n$ limit, in which the Higgs
field is constrained to the vacuum manifold, and suggested
that large skyrmions always tend to expand, because
the magnetic energy, which breaks the approximate
scale-invariance of the string tension, gives smaller skyrmions a
larger energy. Hindmarsh
examined the special case $\beta =1$ and demonstrated the
existence of a one-parameter family of static solutions to the equations
of motion that continuously interpolates between the embedded
vortex solutions and skyrmion configurations of arbitrarily large
size. Hindmarsh also considered how the energy of these exact solutions
for $\beta =1$ varies as $\beta $ is perturbed away from one and concluded that
for $\beta >1$ skyrmions want to expand, and for $\beta <1$ skyrmions
want to contract.

It is interesting to consider what happens when
the global symmetry in a model with semilocal strings
is gauged. For the
$SU(2)_{global}\times U(1)_{local}$ model considered
by Vachaspati and Ach\'ucarro, the result is essentially
the standard electroweak model. The physics of embedded
strings in the Weinberg-Salam electroweak model
was first investigated by Y. Nambu in 1977 \refmark{\nambu },
and more recently by Vachaspati and collaborators %
\refmark{\tvtwo , \james ,\sew }, who explicitly analyzed
the issue of stability. The stability of ``electroweak" strings
depends on $\beta =(\lambda /g^{\prime 2})$ and the ratio $(g/g').$
The region of stability
in the $\beta $-$(g/g')$ plane is a strip where $\beta <1$ and
$g/g'$ is small. In other words, the stability is not
destroyed when $SU(2)$ is gauged very weakly compared to
$U(1).$

These embedded electroweak string are not ``semilocal"
because every direction on the vacuum manifold can
be compensated by the gauge field $A_\mu .$
(We shall call such strings ``completely-gauged" nontopological
strings, as opposed to ``semilocal" nontopological strings.)
The crucial difference is that for completely-gauged
nontopological strings there are no large competing
skyrmions (because the scalar gradients can be completely
compensated by the gauge field),
and the embedded string can break instead of transforming
into a skyrmion, which in essence is just a fattened string.
The ability of completely-gauged nontopological strings
to break makes them even more fragile than their semilocal
counterparts---a fact whose cosmological consequences will
be discussed more.

In this paper, we further investigate the dynamics of
delocalized ``skyrmionic" field configurations.
A heuristic discussion of skyrmion dynamics is first presented
based on an order-of-magnitude variational calculation.
Then a more careful attempt to define skyrmions for $\beta \ne 1$
is presented. The ``size" of a skyrmion is defined
in terms of its magnetic energy, and the lowest energy configuration
of a given size is found.
Then the time evolution of the resulting one-parameter family
of configurations is examined. The effect of the expansion
of the universe is taken into account and found to be significant.
We also discuss the role of localized
embedded strings and delocalized skyrmionic strings in cosmology.
In particular, we discuss the question of
whether a network of localized ``semilocal"
strings would be formed in a cosmological phase transition,
emphasizing how the standard analysis for topologically stable
strings, which inexorably leads to a network described by
the ``scaling" solution,  breaks down. We conclude that the
prospects of forming a network of semilocal strings are rather
bleak.

The organization of the paper is the following. Section 2
presents some background material on embedded string and
skyrmion configurations. Section 3 deals with skyrmion dynamics.
Section 4 discusses the physics of completely-gauged
nontopological strings. Section
5 discusses cosmological implications. Section 6 presents
some concluding remarks.

\chapter{Semilocal Strings and Skyrmions: An Inevitable Competition}

The simplest model with
nontopological ``semilocal" strings has a
complex doublet $\Phi ={\phi _1\choose \phi _2}.$
\refmark{\vacha}
This adds to the local $U(1)$
symmetry a global $SU(2)$ symmetry---hence
the name ``semilocal." The action for this model is
$$
S=\int d^4x\ \left\{ {-1\over 4g^2}
F_{\mu \nu }
F^{\mu \nu }
+
[(\partial _\mu -iA_\mu )\Phi ]^\dagger
[(\partial ^\mu -iA^\mu )\Phi ]
-
{\lambda \over 2}[\Phi ^\dagger \Phi -v^2]^2
\right\} .
\eqn\action$$
The vacuum manifold ${\cal M}=\{ \phi \ \vert \ \phi ^\dagger \phi =v^2\} $
has the topology of $S^3,$ so that
$\pi _1(G/H)=\pi _1(S^3) $ is trivial.
However, an embedded Abrikosov-Nielsen-Olesen solution
still satisfies the equations of motion, with
the form
$$\eqalign{
\Phi (r,\theta )=&\ v\ \cdot
{f(r)e^{i\theta }\choose 0},\cr
{\bf A}(r)=&\ \hat {\bf e}_\theta \ \cdot {a(r)\over r}\cr
}
\eqn\ansatzno $$
where $f(0)=a(0)=0$ and $f(r), a(r)\to 1$ as $r\to \infty .$
The string core
has a characteristic size $r_{vor} \sim v^{-1},$
inside which  $f(r)$ and $g(r)$ deviate significantly from one.
For $r\gg r_{vor}$
the solution becomes pure vacuum, with $\Phi (x)$
at the minimum of the potential and the
covariant gradients and gauge field curvature
vanishing. The boundary condition at
spatial infinity is best understood by foliating the
vacuum manifold $S^3$ using the action
of the gauged $U(1)$ symmetry on $S^3.$
Passing through each point $\Phi _0\in S^3$
is a unique gauge orbit that has the topology of $S^1$
and coincides with a great circle of $S^3.$
It is energetically favorable for $\Phi (x)$ to vary in directions along
these gauge orbits rather than in directions orthogonal to them, for
the gauge field $A_\mu $ can compensate only parallel gradients.
This is why a configuration
with nontrivial winding number in the gauge orbit is not
necessarily unstable to slipping off the gauge orbit.

The triviality of the gauge orbit on the vacuum manifold, however,
implies the existence of competing skyrmionic strings.
Let us fix the fields
at large $R$ so
that $\Phi (x)$ lies entirely within a single gauge orbit.
Let $\Phi (x)$ have a nontrivial winding number, with ${\bf A}(x)$
completely compensating the gradient in $\Phi (x)$ at large
radii. The fact that $\pi _1(G/H)$ is trivial in the semilocal
case indicates that there exist field configurations with this
boundary for which $\Phi (x)$ nowhere deviates from the vacuum
manifold. For the
$SU(1)_{global}\times U(1)_{local}$ model,
an ansatz for configurations of this sort is
$$\eqalign{
\Phi (r,\theta )=&
v\cdot {f(r)e^{i\theta }\choose g(r)}, \cr
A(r)=&\hat e_\theta \ \cdot \ {a(r)\over r}\cr
}
\eqn\ansatzsk
$$
where $f(0)=a(0)=0,$ $\ g(0)=1,$ $f(r)^2+g(r)^2=1$
for all $r,$ and as $r\to \infty ,$ $\ f,a \to 1$
and $g\to 0.$
At large radii, $\Phi (r,\theta )$ lies on a gauge orbit, which
can be thought of as the equator of a three-sphere. As $r$ is decreased,
$\Phi (r,\theta )$ slips off the equator toward one of the poles.
When $r=0,$ the loop has contracted to a point. The resulting configuration
resembles a two-dimensional texture---hence the name ``skyrmion."
Excluding the magnetic energy and restricting
the Higgs field to the vacuum manifold gives an energy for these
configurations that is invariant under scale transformations.
This is because they have a gradient of order $(v/R)$ spread out
over an area $R^2,$ giving an energy per unit
length of order $v^2.$

This reveals the
crucial difference between field configurations
described by
the Abrikosov-Nielsen-Olesen ansatz \ansatzno ~and field configurations
described by the skyrmion ansatz \ansatzsk ~. The size $R$ of a
skyrmion configuration is variable and can be made arbitrarily
large while keeping
the energy bounded from above. This is because
$\Phi (x)$ remains on (or very close to)
the vacuum manifold. By contrast, Abrikosov-Nielsen-Olesen
configurations (with $g(r)=0$) cannot be made arbitrarily large.
There is a potential energy cost of order $\lambda v^4R^2$ associated with
a large Abrikosov-Nielsen-Olesen configuration.

A model that supports semilocal strings
therefore has competing solutions---the Abrikosov-Nielsen-Olesen
solution with localized magnetic flux and localized deviations from the gauge
orbit, and the skyrmion configurations with
delocalized flux and delocalized deviation from the gauge orbit.
The existence of competing configurations has two
important consequences: (1) The stability of the localized
solutions is not guaranteed by topological arguments.
(2) Even if stable, a localized semilocal string can be
unwound with the addition of a finite amount of energy per unit length.
When a semilocal string is bombarded with a sufficient amount
of energy, it will unwind by becoming an expanding skyrmion
solution. The dynamics of this process will be discussed in more
detail in section 3.

The stability of  semilocal strings to small
perturbations
depends on the dimensionless
parameter $\beta =(\lambda /g^2 )=(m_V/m_S)^2.$
$\beta $ is identical to $\kappa $ in superconductivity, which
separates Type-I superconductors $(\kappa <1)$ from
Type -II superconductors $(\kappa >1).$ For $\beta >1$ the semilocal
strings are unstable to small perturbations, but for
$\beta <1$ they are stable against small perturbations
\refmark{\hinda }.
This can be understood
heuristically in the following manner. Consider a
perturbation of the Abrikosov-Nielsen-Olesen solution in which
the Higgs field in the core is pushed away from $\Phi =0$
toward the vacuum manifold. This perturbation gives a decrease
in the potential energy proportional to $\lambda v^4r_{vor}^2$
at the expense of an increase in gradient energy proportional
to $v^2.$ Since $r_{vor}\approx (gv)^{-1},$ the ratio is of
order $\lambda /g^2.$

\chapter{Skyrmion Dynamics}

We now discuss the dynamics of skyrmions. In section 2 we noted that
excluding magnetic energy and restricting the field $\Phi (x)$ to
the vacuum manifold gives a skyrmion energy that is invariant under
dilatations.
Two effects break this scale invariance, for large skyrmions.
First, there is the magnetic energy which scales as $R^{-2}.$ Second,
there is a correction, also of order $R^{-2}$ but of opposite sign,
from the back reaction of the covariant scalar gradient on the
potential, which reduces $|\Phi (x) |$ inside the skyrmion.  The
relative strengths of these effects determine whether initially static
skyrmions tend to expand or contract.

Before discussing the general case, we examine
the special case $\beta =1,$ which lies on the brink of instability
for the Abrikosov-Nielsen-Olesen solution.
Here the two corrections cancel, yielding a continuous family of static
solutions of degenerate energy that interpolate between the
smallest semilocal string and the largest skyrmions.

We see this by exploiting
a remarkable simplification due to
Bogomol'nyi \refmark{\bogo }. After rescaling
fields and spatial coordinates,
the static
energy functional for the action \action ~
depends only on the parameter $\beta :$
$$
E=v^2\int d^2x\ \left\{ {1\over 2}
B^2
+
[(\partial _k -iA_k )\Phi ]^\dagger
[(\partial _k -iA_k )\Phi ]
+
{\beta \over 2}[\Phi ^\dagger \Phi -1]^2
\right\} .
\eqn\energy$$
For finite energy configurations which wind once
around the gauge orbit at spatial infinity,
Bogomol'nyi rewrites \energy ~ as
$$\eqalign{
E/v^2=2\pi  +\int d^2x&\ \Biggl\{ {1\over
2}[(D_j\Phi) + i\epsilon _{jk}(D_k\Phi )]^\dagger [(D_j\Phi) +
i\epsilon _{jl}(D_l\Phi )]\cr &+{1\over 2} (\Phi ^\dagger \Phi -1 +
B)^2 + {(\beta -1)\over 2}[\Phi ^\dagger \Phi -1]^2
\Biggr\}  .\cr }
\eqn\bogoa $$
Note that for
$\beta \ge 1$ the integral is positive definite and gives the lower
bound $E\ge 2\pi v^2  $ . For $\beta >1,$
this bound
cannot be achieved by any particular field configuration, because the
last two quadratic terms give inconsistent equations for $\Phi
^\dagger \Phi .$ However, for $\beta =1$  the bound can be saturated,
and the energy minimized,
by solving the Bogomol'nyi equations
$$
\eqalign{
&D_j\Phi +i\epsilon _{jk}D_k\Phi =0 ,\cr
& B= 1-\Phi ^\dagger \Phi .\cr
}\eqn\bc $$
In polar coordinates, the first equation becomes
$D_{\hat r}\Phi + iD_{\hat \theta }\Phi =0.$  Hindmarsh applies
these equations to
a skyrmion ansatz
$$
\eqalign{
&\Phi =
{f(r)e^{i\theta }\choose g(r)},\cr
\cr
& {\bf A}=\hat {\bf e}_\theta {a(r)\over r},\cr } \eqn\bd $$
with
$f(0)=a(0) = 0, ~g(\infty )=0, ~f(\infty )=a(\infty )=1.$
Like the naive ansatz \ansatzsk , such skyrmions coincide with the
semilocal string at infinity, and unwind through the vacuum
manifold; however, they remain free to slip off the vacuum
manifold in their cores.
For them, the Bogomol'nyi equations \bc ~ become
$$ \eqalign{
&
{d f\over dr}
+{a-1\over r}f=0,
\cr
&
{dg\over dr}
+{a\over r}g=0,\cr
&
B+(f^2-g^2-1)=0.
\cr
}\eqn\be $$
Note that
taking $g = (\Omega/r) f$
eliminates the second equation.
Hindmarsh showed that
for all values $\Omega \in [0,\infty )$
there exist solutions to
\be ~  satisfying the boundary conditions.
When $\Omega =0$,
the solution has $g \equiv 0$ and therefore coincides with
the embedded Abrikosov-Nielsen-Olesen solution.
When $\Omega $ increases, $g(r=0)$ increases and the region
over which the solution's energy is spread becomes larger.
For $\Omega \gg 1,$ $\Phi ^\dagger \Phi \approx 1$  everywhere,
and the solution is essentially a two-dimensional texture.
Therefore , for $\beta =1$ skyrmions
interpolate between the semilocal string and
skyrmion solutions of arbitrary size.

We now consider $\beta \ne 1,$ where we no longer expect skyrmions to
be static solutions. Rather they are configurations of fixed size,
which tend to expand or contract. Such objects comprise a trough in
configuration space: at $\beta = 1$, their tension is scale-invariant
and the trough is flat; however, we generically expect corrections to
tilt the trough, biasing it toward either large or small
skyrmions.

To understand this bias we must
relate the skyrmion energy $E$ to its size $R.$
By $R$ we mean the size of the
region in which the gauge curvature and the covariant derivative of
the scalar field differ significantly from zero.
The skyrmion has scalar derivative energy of order $v^2$, because a
covariant gradient of order $(v/R)$ extends over an area of order
$R^2$. It has a smaller magnetic energy of order $1/g^2R^2$.  A
further correction appears when we refine the approximation for the
gradient energy to take into account the scalar field's freedom to
move off the vacuum manifold in the skyrmion core. Suppose that inside
the region of size $R,$ $\ \Phi ^\dagger \Phi \approx x^2v^2 $ where
$x<1.$ This lowers the energy by
lowering the gradient energy linearly in
$(x^2-1)$ at a cost in potential energy quadratic in $(x^2-1).$
The energy then takes the form $$ E(R,x)=
\left(
2\pi{v^2x^2\over R^2}
+{b\over g^2R^4}
+c\lambda v^4(x^2-1)^2
\right)\cdot R^2
\eqn\ma
$$
where $b$ and $c$ are undetermined coefficients
of order one. Minimizing with respect to $x$ for fixed $R$ gives
(to leading order in $R^{-2})$
$$E(R)=2\pi v^2+\left( {b\over g^2}-{\pi\over 2c\lambda } \right)
{1\over R^2}.
\eqn\mba$$
A relation between the coefficients $b$ and $c$ can be
found using the special case $\beta =1.$ The fact that the $R^{-2}$
coefficient vanishes for $\lambda =g^2$
allows one to simplify \mba ~ to
$$E(R)= 2\pi v^2+b\cdot \left( {1\over g^2}- {1\over \lambda }\right) %
{1\over R^2}.\eqn\md$$

This order of magnitude analysis
can be refined by seeking the skyrmions configurations explicitly.
They are minimal energy configurations of fixed size, whose
spectrum biases them toward shrinking or growing when $\beta \ne 1.$
Finding them thus reduces to the question of minimizing the
energy for fixed skyrmion size. There
are many possible definitions for this ``size"---all of which agree
roughly, but not exactly---with no way to single out a particular
definition as the most natural. We therefore define the size to make
the calculation simple.

We define a skyrmion's ``size'' as its ratio of flux energy to
total energy. To confine a skyrmion configuration to a smaller region,
we pay a price in flux energy, as we concentrate the flux lines.
For large skyrmions, this cost scales strongly, $ E_{mag}\sim 1/R^2,$
while the total energy $E$ changes negligibly.
Our parameter $\Chi =E_{mag}/E$ thus acts as an ``antisize.''

To find a minimal energy configuration for fixed $\Chi$,
we introduce a Lagrange multiplier to constrain $\Chi$ and minimize
$$ \bar E=E-\alpha[E_{mag}-\Chi E],
\eqn\ka $$
which can be rewritten as
$$
\eqalign{
\bar E=&(1+\alpha \Chi )E^\prime \cr
=&(1+\alpha \Chi ) v^2
\int d^2x\ \left\{
\vert D_i\Phi \vert ^2+
{\beta \over 2}[\Phi ^\dagger \Phi -1]^2
+{1\over 2}\left(
{1+\alpha \Chi -\alpha \over 1+\alpha \Chi }
\right) B^2 \right\} .\cr }
\eqn\kb
$$
In this form we see the result of constraining $\Chi$: it acts
only to rescale the flux energy contribution $B^2$. We trade this
rescaling for a rescaling of $\beta$, by rescaling
$ x\to x'=\left( {1+\alpha \Chi \over 1+
\alpha \Chi  -\alpha }\right)^{1/2} x,$
which does not affect the energy.  This eliminates the
factor $\left( {1+\alpha \Chi -\alpha \over 1+\alpha \Chi }\right) $
in front of $B^2$, and transforms $\beta $ to
$$ \beta _{eff}=\beta
\left( { 1+\alpha \Chi -\alpha \over 1+\alpha \Chi }
\right) .
\eqn\kc $$
This makes the problem tractable because we know how to
minimize the energy
when $\beta _{eff}=1.$ Thus setting
$$
\alpha ={\beta -1\over \beta -\Chi (\beta -1)}
\eqn\kd $$
gives solutions to the constrained problem
$$ \eqalign{f_\beta (r;\Omega )&= f_{1}(r';\Omega )\cr
g_\beta (r;\Omega )&= g_{1}(r';\Omega )\cr
a_\beta (r;\Omega )&= a_{1}(r';\Omega )\ ,\cr }
\eqn\ke $$
where $r'=\left( {1+\alpha \Chi \over 1+\alpha \Chi %
-\alpha }\right)^{1/2} x =\sqrt{\beta } r$ and $f_{1}(r';\Omega ),\ldots $ are
the solutions for $\beta =1$. This yields a
one-parameter family of skyrmions, ranging from infinite size to a
smallest member at $\Omega = 0$. Since this smallest member is just
a rescaling of the $\beta = 1$ semilocal string, it
has a particle-scale size of order $(1/\sqrt{\beta})v^{-1}.$

Because these skyrmions have $E' = 2\pi v^2$, equation \kb ~gives their
spectrum exactly:
$$E(\beta ,\Chi )=2\pi v^2\cdot (1+\alpha \Chi )=
2\pi v^2\cdot \left( {\beta \over \beta -\Chi (\beta -1)}\right) .
\eqn\kf $$
This function resolves qualitative questions about the skyrmions' flat
space dynamics. $E(\Chi)$ increases monotonically when $\beta > 1$,
and decreases monotonically when $\beta < 1$. Thus skyrmions do tend
to grow (to lower their flux fraction $\Chi$) when $\beta > 1$,
and shrink when $\beta < 1$.  Moreover, no energy barriers arise in
this evolution; the skyrmions evolve between one endpoint---an
infinite skyrmion---and the other---a particle-scale skyrmion---%
classically and continuously.
Over all these orders of magnitude, the energy \kf
varies over at most the range
$$
E: 2\pi v^2 \to 2\pi v^2 \beta .
\eqn\range$$
Thus the finite energy bias
$$
\Delta < 2\pi v^2 |\beta -1|
\eqn\bias$$
drives skyrmion expansion or contraction from cosmological to particle
scales.

Note that the rescaled $\beta =1$ solutions do not solve the
constrained problem for all values of $\Chi .$
Values of $\Chi $ greater than $\beta /(1+\beta )$
are inconsistent with the Bogomol'nyi equations \be .
For $\beta \ne 1$ the Lagrange multiplier $\alpha $
remains nonzero for all of the rescaled $\beta =1$
solutions. This means that the rescaled $\beta =1$
solutions never reach the size of the localized
Abrikosov-Nielsen-Oleson solution for $\beta \ne 1.$

With these energetic tendencies thus established, we
address the
dynamics of cylindrically-symmetric skyrmions, described by a function
$R(t),$ in flat Minkowski space.
The Hamiltonian for the restricted degree of freedom $R(t)$
contains an inertia term of the form ${1\over 2}Iv^2\dot R^2(t),$
corresponding to the gradient $(\partial \Phi /\partial t)\approx
(v/R)\dot R$ spread over an area of order $R^2.$ $I=O(1).$ Large
skyrmions have a potential obtained by expanding the spectrum \kf ~in
powers of $\Chi$:
$$
E =  2 \pi v^2\left ( 1 + \Chi {\beta -1 \over\beta }  + O(\Chi^2) \right ) \ .
\eqn\bigE$$
Given the large skyrmion relation $\Chi \sim v^{-2}/R^2,$ this agrees with
the order of magnitude spectrum \md .
The full Hamiltonian is
$${\cal H}={1\over 2}Iv^2\dot R^2+ 2\pi v^2 -
{b\over\lambda}\cdot \left( 1- \beta \right) {1\over R^2} .
\eqn\me$$
leading to the equation of motion
$$\ddot R=-{2b\over \lambda I v^2}
\cdot \left( 1- \beta \right) {1\over R^3}
={-\zeta \over v^2R^3}.
\eqn\mf$$
Thus, when
 $\beta < 1$ skyrmions
always shrink, at a rate
$$
\dot{R } \sim - {1\over \sqrt{1 - \beta }} \ \ {1 \over v R}
\eqn\contract$$
which results in the quadratic collapse time
$$
t \sim \sqrt{{1\over 1-\beta}} \ \ v R^2 \ .
\eqn\colltime$$

However, in the early universe, the skyrmion also experiences the pull
of the universe's expansion, so that for a flat skyrmion potential
$$
\dot{R} \sim H R \ ,
\eqn\ja$$
where $H$ is the Hubble constant.  The conflicting effects determine a
cutoff size
$$
R_c \sim \sqrt{(Hv)^{-1}}\ ,
\eqn\cutoff$$
which is the geometric mean of the two length scales in the problem:
the localized string core and the horizon size.  For $R \gg R_c$, the
expansion wins, and the skyrmion expands with the universe. For
$R \ll R_c,$ the skyrmion evolves instead as predicted by microphysics,
collapsing slowly toward the semilocal string. Note that, in the
radiation-dominated universe, the cutoff $R_c$ between these two
regimes itself comoves with the expansion,
$$
R_c = \sqrt {{M_{pl}\over v}} \ \ T^{-1}.
\eqn\jb$$

This cutoff is approximate,
and the large expanding skyrmions do not quite
comove with the universe, due to their retarding dynamics. Thus some
skyrmions which initially expand are recaptured: they fall within the
comoving cutoff $R_c$ and succumb to dynamical collapse. However, we
show below that this effect is small, leading to recapture only of
skyrmions with initial sizes within an order of magnitude of $R_c$.
Larger skyrmions expand forever, since their dynamical drag becomes
ineffectual after matter-radiation equality.  Before that refined
analysis, however, we illustrate the relevance of the cutoff $R_c$
with some numbers.

A phase transition at the GUT scale ($10
^{16}$ GeV) yields a semilocal string which spans $v/M_{pl} = 10^{-3}$
of the initial horizon.  Thus skyrmions smaller than roughly $1/30$ of
the horizon shrink; larger ones expand.  (Note that this cutoff $R_c$, a
geometric mean, also corresponds to 30
times the semilocal core size $v^{-1}$.)
To forbid initially expanding skyrmions,
the correlation length must be kept
smaller than $1/30$ of the horizon size.  Folding in our promised
result, that less than one order of magnitude of expanding skyrmions
recollapse, this bound loosens. Skyrmions up to $1/3$ of the horizon
size eventually collapse into semilocal string.

While this poses no restriction for GUT scale transitions, it becomes
problematic as we lower $T_c$.  A phase transition occurring $13$
orders of magnitude later, at the electroweak scale ($1$ TeV), yields
semilocal strings which span only $v/M_{pl} = 10^{-16}$ of the initial
horizon. Thus skyrmions bigger than $10^{-8}$ times the horizon size
(or $10^8$ times the semilocal core size) begin expanding.  Again
anticipating our promised result, only one order of magnitude of these
expanding skyrmions recollapse. Thus, to prevent skyrmions which
expand forever, we must require a correlation length less than
$10^{-7}$ horizon sizes --- a significant fine tuning of the phase
transition.

We now proceed to the promised result, a careful analysis of skyrmion
dynamics in an expanding FRW universe with $\Omega =1.$ First, it is
convenient to rewrite our equation of motion \mf ~ in terms
of the dimensionless co-moving size $x(t)=R(t)/a(t)$ (where
$a(t)$ is the scale factor) instead of the physical size
$R(t).$ One obtains
$$ \ddot x(t)+{\zeta \over v^2a^4(t)x^3(t)}=0,\eqn\za$$
valid for the static case $\dot a(t)=0.$ To generalize, we
substitute
$\ddot x \to \ddot x +3[\dot a(t)/a(t)]\dot x(t),$ to obtain the curved
space equation of motion
$$ \ddot x(t) +{3\dot a(t)\over a(t)}\dot x(t)
+{\zeta \over v^2a^4(t)x^3(t)}=0.\eqn\zb$$
For a radiation-dominated universe (with a fixed effective
number of massless species),
$a(t)=t_0\cdot [t/t_0]^{1/2},$ where $t_0$ is an arbitrary time
scale used to fix a normalization for $x(t).$ Here $H(t)=1/(2t).$
With this choice of $a(t),$ \zb ~ becomes
$$ \ddot x(t) +{3\over 2t}\dot x(t)
+{\zeta \over v^2t_0^2}\left({t_0\over t}\right) ^2 {1\over x^3(t)}=0.
\eqn\zb$$
Rewritten in terms of logarithmic time $s={\rm ln}[t/t_0],$
\zb ~ becomes
$$
{d^2x\over ds^2}+{1\over 2}{dx\over ds}+
{\zeta \over (vt_0)^2}\cdot {1\over x^3(s)}=0.\eqn\zc$$

At this point, we normalize
$x(t)$ so that at the phase transition $t=t_{pt},$
$x=1$ corresponds to the physical size
of the Abrikosov-Nielsen-Olesen vortex---that is,
$a(t_{pt})\approx v^{-1}.$ This occurs for
$t_0=M_{pl}^{-1},$ so that \zc ~ becomes
$$ {d^2x\over ds^2}+{1\over 2}{dx\over ds}+
\zeta\left( {M_{pl}\over v}\right) ^2{1\over x^3(s)}=0.\eqn\zd$$
Note this equation is invariant to translation in logarithmic time $s$;
we thus set $s(t=t_{pt}) = 0$.
We want to know at what later value of $s$
a skyrmion initially at rest with size $x$ completes its collapse. Since
the final stages of collapse occur rapidly, it is justified
to assume that the collapse completes when $x=0.$

Note that equation \zd ~describes dynamics in the ``time'' $s$, where
the force acts always to push the velocity and acceleration (both
initially zero) in the same direction. Thus a positive acceleration
always reduces the velocity from its terminal value,
$$ {1\over 2}{dx\over ds}+
\zeta\cdot \left( {M_{pl}\over v}\right) ^2\cdot {1\over x^3(s)}=0.\eqn\ze$$
Using this terminal velocity, then,
we underestimate the collapse time. We obtain
$$ s_{col}(x) = \int_x^0 \left ( {dx \over ds} \right ) ^{-1} dx
\gtorder {1\over 8\zeta}\cdot
\left( {v\over M_{pl}}\right) ^2 x^4.\eqn\zf$$
Recall that $t_{col}=t_{pt}\cdot e^s,$
so rather modest values of $s$ can represent exceedingly long times.
We see this by calculating the size of the smallest skyrmions which fail to
 collapse before recombination $(T _{rec}\approx 1eV).$
Such skyrmions have collapse times larger than
$s_{max}=
{\rm ln}[t_{pt}/t_{rec}]
\approx {\rm ln}[v^2/T_{rec}^2].$
This bounds their size
$$
x\gtorder x_{c}\approx \left( {M_{pl}\over v}\right) ^{1/2}
\cdot \bigl( 8\ \zeta \ s_{max} \bigr) ^{1/4} .\eqn\zh$$
We compare this with our first order of magnitude analysis.  Equation
\jb ~ gives the naive physical cutoff $R_c \sim \sqrt {M_pl/v} \ v^{-1}$
at the phase transition. From our normalization $a(t_{pt}) = v^{-1}$.
Thus this more careful analysis modifies the naive
cutoff as follows:
$$
x_{c}\approx x_{c, naive}
\cdot \bigl( 8\ \zeta \ s_{max} \bigr) ^{1/4}
\approx x_{c, naive}
\cdot 2 \left ( {\rm ln} v/T_{rec} \right )^{1/4}
.\eqn\zi$$
The correcting factor accounts for the fact that initially expanding
skyrmions eventually recollapse, over the cosmological time scales we
consider here. Given the length of those time scales, the correction is
amazingly weak. For a GUT scale phase transition, $v/T_{rec} =
10^{25}$ gives only $x_{c} \sim 6 x_{c, naive}$. That is, only
skyrmions $6$ times bigger than those which begin collapsing instantly
are ever recaptured, even as we evolve through $25$ orders of
magnitude in temperature. For an electroweak phase transition, things
change little: $v/T_{rec} = 10^{12}$ gives $x_{c} \sim 5 x_{c,
naive}$.

For a matter-dominated universe, the conditions for collapse
become even less favorable, because the expansion of the universe
slows less rapidly. In a matter-dominated universe,
$a(t)=t_0\cdot [t/t_0]^{2/3},$ so \zb ~ becomes
$$\ddot x(t)+{2\over t}\dot x(t)+{\zeta \over v^2t_0^4}\cdot
\left( {t_0\over t} \right) ^{8/3}\cdot {1\over x^3(t)}=0.\eqn\zm $$
Again using logarithmic time $s={\rm ln}[t/t_0],$ one obtains
$$
{d^2x\over ds^2}+{1\over 2}{dx\over ds}+
{\zeta \over (vt_0)^2}\cdot e^{-(2/3)s}{1\over x^3(s)}=0.
\eqn\zn$$
The crucial difference between \zc ~
[radiation domination with $p\approx \rho $] and \zn ~
[matter domination with $p\ll \rho $] is the presence
of the factor $e^{-(2/3)s},$ which causes the potential term
in the matter-dominated universe to become feeble exponentially,
so that large skyrmions never collapse, even in the $t\to \infty $
limit. Therefore, one expects most skyrmions that survive till
matter domination to survive forever without collapsing.

\chapter{Completely-Gauged Nontopological Strings}

As mentioned in the Introduction, the
stability of embedded strings
 does not rely on the existence of global symmetries.
Instead, the global symmetries can be weakly
 gauged (by introducing new gauge
fields whose couplings are
 much smaller than the pre-existing ones),
without destroying the stability of semilocal
solutions. However, making the gauge couplings
roughly comparable does destroy their stability.
Vachaspati and collaborators studied
the model \action ~ with an $SU(2)$
gauge field added (i.e., the Standard Electroweak Model) and
found embedded $Z$-strings which are stable for values of
$\sin \theta _W$ close to one \refmark{\tvtwo ,\james ,\sew }.
These strings carry mostly
$U(1)$ flux, with a small admixture of
 $SU(2)$ flux, which is energetically disfavored because
$g'\gg g.$ (Here $g$ is the $SU(2)$ gauge coupling and $g'$ the
$U(1)$ gauge coupling.) To deform such an electroweak vortex to a
configuration
that is pure gauge, one must first turn the $U(1)$ flux into
energetically more costly $SU(2)$ flux, which involves climbing
over an energy barrier of order $1/(g^2r_{vor}^2).$

The most important consequence of gauging the global symmetry is the
absence of large skyrmions.  Large skyrmions arise only in embedded
models that are ``partially gauged"---that is, models which have
directions on the vacuum manifold that cannot be compensated by gauge
fields. In this case, unwinding the gauge orbit at spatial infinity
--- by contracting it in the vacuum manifold --- always has some
gradient energy cost. To first order, that cost is scale-invariant and
of order $v^2$. It is borne most efficiently by the skyrmions, or
textured string configurations.

In models that are ``completely gauged," (such as the electroweak
model with $g\ne 0$), every direction on the vacuum manifold can be
compensated by gauge fields. Thus, when we contract the gauge orbit in
the vacuum manifold, we can offset all gradients by exciting the
weakly coupled gauge field. This carries a cost in flux energy of
order $1/g^2 R^2$ --- expensive compared to $U(1)$ flux, but
negligible compared to the finite gradient energy of skyrmions. Thus
we see the key difference between semilocal and completely gauged
theories: in semilocal theories, even infinitely delocalized
configurations with $U(1)$ winding carry an energy of order $v^2$;
whereas, in completely gauged theories, configurations with such
delocalized winding can cost negligible energy.

Note that these delocalized configurations, in the completely gauged
case, invariably expand, driven by their flux energy. Thus completely
gauged embedded strings get only one opportunity in the competition to
resolve Kibble frustration. They must form at the phase transition, or
not at all, as their negligible-energy competitors expand forever
once they form.

The absence of large skyrmions also implies that completely-gauged
nontopological strings can break, \refmark{\pv }
without forming a fat
skyrmionic string of comparable or often larger energy
to fill the gap.
As Nambu explained,
this does not mean that electroweak strings can end into
nothingness \refmark{\nambu }.
The string carries a $U(1)_Y$ magnetic hypercharge
and $U(1)$ lines of magnetic induction must close on themselves.
This means that an ``electroweak" string ends on something
resembling a magnetic monopole. Inside the string there is
$Z$-magnetic flux, which is a mixture of $SU(2)$ magnetic flux and
$U(1)_Y$ magnetic flux. Where the string ends, this $Z$-magnetic flux
turns into $Q$-magnetic flux in such a way that $Y$-magnetic flux
is conserved. The end of the string is not really a magnetic monopole
because the Dirac string is physical and the energy density at the
end of the string is comparable to the energy density
elsewhere in the core of the string. Therefore, in order not
to collapse,
a finite segment of electroweak string must rotate so that the
velocity at the end of the string is approximately the
velocity of light.

There are also models with completely-gauged nontopological
strings that can end into nothingness. Vachaspati and Barriola
\refmark{\mb } found a ``W"-string solution that can be embedded
in the Standard model. In this solution the $U(1)_Y$ gauge
field is not excited; therefore, the string is free to break
without any magnetic flux filling the gap.

\chapter{Semilocal Strings and Cosmology}

We now discuss the formation and evolution
of a network of nontopological string in the early universe.
We conclude that the usual analysis of string network formation,
valid for topological
cosmic strings, cannot naively be extended to nontopological strings.
These are both more difficult to form and
more fragile after formation than their topological counterparts.

For topological strings, the standard
picture for formation is the following. \refmark{\kibble ,\zel ,\vila }
As the universe cools, a
phase transition takes place and
the Higgs field $\Phi (x)$ acquires a nontrivial
vacuum expectation value, with the
orientation of $\Phi (x)$ uncorrelated over length scales larger than
the correlation length $\xi .$
This correlation length $\xi$ corresponds for
second-order phase transitions to the length scale where critical
slowing down prevents $\xi$ from further growth and
for first-order phase
transitions to the separation between bubbles.
In either case, causality bounds $\xi$ to be smaller than
the Hubble length $H^{-1}.$

Typically, to simulate string
formation in this process, one takes a lattice with spacing $\xi $
and lays down random uncorrelated orientations of the Higgs field at the sites
of the lattice \refmark{\vv }.
A rule is established for interpolating $\Phi (x)$
along the links connecting adjacent sites, so that the winding number $N$
throug
 h each
plaquette can be calculated. For example, for the $U(1)$ abelian Higgs model,
a three-dimensional triangular lattice can be used, with $\Phi (x)$
randomly set to $1,\  e^{i2\pi /3},\  e^{i4\pi /3}$ (which we shall denote as
$1, 2, 3,$ respectively) at each site.
$123$ around a plaquette (and its two cyclic permutations) corresponds to a
wind
 ing number
$N=+1,$ $213$ (and its cyclic permutations) corresponds to $N=-1,$
and the other 21 possibilities correspond to $N=0.$

For topological strings, the winding number $N$ through each plaquette
is equal to the number of strings minus the number of antistrings
passing through the plaquette.  For the $U(1)$ example each
tetrahedral simplex has either no strings passing through it, or one
string entering and leaving the simplex. Therefore, there is no
ambiguity in determining the trajectories of the strings from the
winding numbers on the plaquettes. The end result is a scale-invariant
distribution of closed loops (dominated by small loops) and some open
infinite strings. \refmark{\vv ,\preskill }

Unlike theories with topological strings, ``semilocal'' systems can
accommodate winding boundary conditions through a multiplicity of
low-energy configurations.
They can untwist internally by producing a tightly
localized semilocal core, or a skyrmionic core with size up to the
correlation length.  Thus, for nontopological strings, the winding
number $N$ does not determine the number of ``strings'' passing
through a plaquette.
This requires more information than simply the
behavior of the fields on the boundary of the surface. When $\xi $ is
much larger than $v^{-1}$ ( the thickness of the
semilocal string), there is a high probability that
$N\ne 0$ will correspond to a large delocalized skyrmion, and not to a
localized string.

The distinction between ``localized" and ``delocalized" is
not a sharp one. On the one hand, if frustration resolves
on scales larger than $v^{-1}$ but at the same time smaller
than the length scale $L_f$ characterizing the frustration density
(so that voids may appear between the fattened skyrmionic strings),
the result is best described as a network of fattened string.
But, on the other hand, if skyrmions have a thickness comparable
to $L_f,$ then perturbations in energy density---at least
initially---will hardly resemble that of a string network.
Because there are no voids (where the energy density perturbation is
negligible), the skyrmions will more closely resemble a global texture.
While it is reasonable in former case to count skyrmions
as ``strings," in the latter case to do so would be misleading.

For topological strings, the lattice model---as crude as
it is---guarantees that there are at least some open
infinite strings.
As demonstrated by simulations and analytical models,
an initial string distribution with some open infinite strings
rapidly settles down to what
is known as the ``scaling solution" \refmark{\vila ,\ktwo },
in which there is approximately
one infinite string per horizon volume
and a scale-invariant distribution
of large string loops.
The expansion of the universe tends to stretch and straighten out
long cosmic strings. If cosmic strings did not intercommute, but instead
merely passed through each other upon colliding,
the number of strings per
horizon volume would rapidly grow, and cosmic strings would eventually
dominate the mass density of the universe. But instead
abelian cosmic strings always intercommute upon colliding.
\refmark{\shellard , \matzner ,\mor }
Intercommuting reduces the amount
of infinite string per unit co-moving volume by allowing loops to
be chopped off and by preventing infinite
strings from becoming straightened out
and stretched. Causality prevents the density of infinite strings from
dropping below approximately one infinite string per horizon volume.
The scaling solution is stable in the sense that any
initial distribution
of cosmic strings that contains some infinite strings will approach
the scaling solution as $t\to \infty .$

The essential input required for a distribution of cosmic strings to
approach the scaling solution is that there initially be a
nonvanishing density of infinite strings.
There is no compelling
reason to believe that any open, infinite semilocal strings are
generically produced at the phase transition.  If one
assumes that a frustrated plaquette has some probability $p<1$ of
corresponding to a semilocal string and a probability $(1-p)$ of
corresponding to a large skyrmion, then one will obtain closed loops
of semilocal string and open segments of semilocal string that connect
to skyrmions on both ends. The distribution of closed loops initially
will be exponentially suppressed from a $dn/dR\sim R^{-4}$
scale-invariant distribution, and there will be no infinite, open
semilocal strings---at least immediately after the phase transition.

Subsequently, for $\beta <1,$ finite segments of semilocal string
lengthen, and some of the skyrmionic string starts to collapse. But
these two mechanisms cannot produce infinite semilocal strings unless
the correlation length is restricted to small values,
less than approximately $10\cdot (M_{pl}/v)^{1/2}v^{-1}.$
Larger correlation lengths give rise to
delocalized skyrmionic string, which spreads with the Hubble flow and
expands forever. Finite-length segments of semilocal string whose flux
flows into such large, forever expanding regions of delocalized flux
effectively stop their growth, because their small energetic advantage
must act against a huge and increasing inertia.  Even smaller
skyrmions, which begin to collapse, offer no guarantee of eventual
semilocal string. Their delocalized flux and frustration can also
relax by interacting and subsequently annihilating with nearby
delocalized flux and frustration. In other words, skyrmions can
disappear without ever becoming localized.

We have yet to consider thermal effects, which destabilize the
semilocal string for $\beta $ close to one.  Immediately after the
phase transition, the temperature is of order $v,$ but as $\beta \to
1$ from below the energy gap between semilocal strings and delocalized
skyrmions becomes negligible compared to $v$. Thus one expects thermal
bombardment to destabilize any short segments of semilocal string
formed near the phase transition.

For ``completely-gauged" nontopological strings, the prospects for
formation are even more bleak
because there are no competing large skyrmion configurations,
with energy per unit length of order $v^2.$
Completely-gauged nontopological strings that do not carry any abelian
magnetic flux can simply end into
pure vacuum. This fact implies that even if such strings are classically
stable (that is, stable against small perturbations), they
break at a finite rate due to quantum tunneling processes. The
Euclidean action for the instanton for this process is not
very large because there is no need to create heavy monopoles.
\refmark{\pv }. Thermal fluctuations also break completely-gauged
nontopological strings.  When completely-gauged
nontopological strings carry some abelian
magnetic flux, the situation differs in one minor respect.
The string cannot simply end because $\nabla \cdot {\bf B}=0$ for an
abelian magnetic field. This means that where the string breaks,
there must be magnetic flux in the gap connecting the two string
segments. However, the energy between the two strings is minuscule
because there is no scalar gradient energy. Therefore, energy
per unit length in a gap of width $R$ is roughly $1/(g^\prime R)^2.$

Completely-gauged nontopological strings are unlikely to form, except
perhaps for some small loops, because the competing large configurations
have completely negligible energy.
Instead of competing with skyrmions of comparable tension,
biased to collapse into localized string, they compete with delocalized
configurations of negligible tension which tend to expand. After the
phase transition there are no delocalized strings
lurking around with the
potential of collapsing into localized strings. Localized strings
either form immediately after the phase transition or not at all.

\chapter{Concluding Remarks}

Nontopological embedded string solutions are more fragile than
their topologically stable counterparts. A nontopological string can
be unwound by supplying a finite energy per unit length.  If the
nontopological string is ``semilocal,"  the string unwinds by
turning into a fat ``skyrmionic" string whose radius increases at an
asymptotically constant velocity. If the nontopological string is
``completely gauged," it can disappear, merely turning into
outgoing particles and radiation.  By contrast, it requires an energy
proportional to $R$ to delocalize the winding of a
topologically stable string to a cylinder of radius $R.$ Therefore,
an infinite amount of energy per unit length is required to unwind a
topologically-stable string.

In a cosmological context, the presence of stable nontopological
strings in a model does not mean that a string network, described by
the scaling solution, invariably forms due to the Kibble mechanism.
Formation of such a network can occur only for correlation lengths
$\xi \ltorder 10\cdot \sqrt{M_{pl}/v}\cdot v^{-1}$
for semilocal strings, and $\xi \ltorder (1/g)v^{-1}$ for
completely-gauged strings. Furthermore, even if a completely-gauged
string network forms initially, it is later destroyed by thermal
and instanton effects.

However, even if a string network does not form in a semilocal model
owing to the role of expanding
delocalized ``skyrmions,"
this does not mean that the resulting skyrmions are
cosmologically uninteresting. Density perturbations will still be
produced and could be relevant to structure formation;
however, the density perturbations will not be ``stringy"---
that is, they will not have a filament-like structure.

In this paper we have studied the dynamics of delocalized skyrmion
configurations, both in a static universe and in an spatially flat
expanding universe. We found that for large skyrmions the
potential scales as $E(R)\sim (\beta -1) R^{-2},$ becoming exceedingly
weak for large $R.$ Therefore, the
expansion of the universe becomes relevant---and even dominant---for
$R$ substantially smaller than the horizon size.
Skyrmions of a rather modest size are swept away by the
Hubble flow and never collapse, despite the fact that their collapse
would lead to lower
energy.  We consider the relevance of these competing fat skyrmionic
strings to cosmology, demonstrating that they prevent the extension of
the standard picture for string formation, valid for
topologically stable strings, to semilocal strings.  One of the
disappointing results of this investigation is the fact that the
formation and evolution of ``nontopological" strings depend crucially
on the length scales which arise at the phase transition. By contrast,
for ``topological" strings (and other types of topologically-stable
defects), the details of the phase transition become largely
irrelevant at later times. For topologically-stable strings, the fact
that the details of the phase transition are quickly erased leads to
more generic, model independent predictions.

\vskip 15pt
\leftline{\bf Acknowledgments}

We would like to acknowledge helpful discussions with
Andrew Liddle, John Preskill, Tanmay Vachaspati,
Alexander Vilenkin, David Weinberg, and Frank Wilczek.

\refout
\end